\documentclass[aps,prd,reprint,groupedaddress]{revtex4-1}

\usepackage{graphicx}
\usepackage{booktabs}
\usepackage{amssymb,bm,mathrsfs,bbm,amscd}
\usepackage[tbtags]{amsmath}
\usepackage{lastpage}
\usepackage{CJK}

\begin{document}


\title{Effects of the tensor couplings on the nucleonic direct URCA processes in neutron star matter}
\thanks{Supported by Natural Science
Foundation of China under grants Nos.11447165, 11373047}



\author{Xu Yan$^{1}$}
 \email{xuy@cho.ac.cn}
\author{Huang~Xiu-Lin$^{ 1,2 }$}%
 \email{huangxl10@mails.jlu.edu.cn}

\author{Liu~Cheng-Zhi$^{1}$}%
 \email{lcz@cho.ac.cn}

\author{Liu~Guang-Zhou$^{2}$}

\address{%
$^1$  Changchun Observatory, National Astronomical Observatories,
Chinese Academy of Sciences, Changchun 130117, China\\
$^2$  Center for Theoretical Physics, Jilin University, Changchun 130012, China}



\begin{abstract}
The relativistic neutrino emissivity of the nucleonic direct URCA processes in neutron star matter are investigated within the relativistic Hartree-Fock approximation.
We study particularly the influences of the tensor couplings of vector mesons $\omega$ and $\rho$ on the nucleonic direct URCA processes.
It is found that the inclusion of the tensor couplings of vector mesons $\omega$ and $\rho$ can slightly increase the maximum mass of neutron stars.
In addition, the results indicate that the tensor couplings of vector mesons $\omega$ and $\rho$ lead to obvious enhancement of the total neutrino emissivity for the nucleonic direct URCA processes, which must accelerate the cooling rate of the non-superfluid neutron star matter.
However, when considering only the tensor coupling of vector meson $\rho$, the neutrino emissivity for the nucleonic direct URCA processes slightly declines at low densities and significantly increases at high densities.
That is to say that the tensor coupling of vector meson $\rho$ leads to the slow cooling rate of a low-mass neutron star and rapid cooling rate of a massive neutron star.
\end{abstract}

\pacs{97.60.Jd, 26.60.Dd, 95.30.Cq}

\maketitle


As you know, the nature of an neutron star (NS) is closely related to nuclear physics, particle physics and astrophysics.
Specifically, the recent measurements of millisecond PSR J 1614-2230 extend the maximum observed mass from $1.67\pm0.02M_{\odot}$ to $1.97\pm0.04M_{\odot}$\cite{Demorest2010}.
Therefore, the study on the properties of NS is one of the hottest issues in the field of nuclear astrophysics.
NSs are the remnants of supernova explosions with an internal temperature about $10^{11}$-$10^{12}$ K, and reduce to temperature about $10^{10}$ K within minutes by emitting neutrinos\cite{Shapiro1983White}, then NSs reach the steady thermal states.
The long-term cooling process of an NS is mainly by the neutrino energy losses in the NS interior and can keep about $10^{6}$ years.
During this period, the neutrino emissions can be classified into two categories:  the enhanced neutrino processes that mainly mean the nucleonic and hyperonic direct URCA processes\cite{lattimer1991direct,levenfish1994suppression,yan2013nucleon,xu2014direct}, and the standard neutrino processes(such as modified URCA processes and Bremsstrahlung processes).
It is well known that the nucleonic direct URCA processes produce the most powerful
neutrino emissivity. They dominate the properties of NS cooling and are more efficient than standard processes\cite{Prakash1992Rapid,Haensel1994Direct,Yakovlev2000Neutrino}.
In recent years, the tensor couplings of vector mesons $\omega$ and $\rho$ have been used in studies of nuclear properties with effective Lagrangians and results have shown that they can significantly impact on the overall nuclear properties\cite{Furnstahl1998The,Mao2002Effect,Alberto2005Tensor,Jiang2005Charge}.
Theoretically, the tensor couplings of $\omega$ and $\rho$ mesons terms in the Lagrangian are allowed by lorentz, gauge, and CP invariance.
NSs belong in the typical nuclear matter.
So it is necessary to study the effects of the tensor couplings of $\omega$ and $\rho$  mesons on the bulk properties of NSs.
It is for exactly these reasons that the relativistic Hartree-Fock (RHF) theory has been used widely in the study of the properties of NSs\cite{Bouyssy1987Relativistic,Huber1996Relativisitic,H2012Neutron,Katayama2012EQUATION,Miyatsu2013A,Sun2008Neutron,Miyatsu2011Effects}.
In the model, the Fock terms are composed of the pion exchange and the tensor couplings of $\omega$ and $\rho$  mesons.
When the tensor couplings of $\omega$ and $\rho$  mesons are included in the RHF approximation, the properties of NSs must be changed as the equation of states (EOSs) of NSs change.
In this work, we adopt the RHF theory including the tensor couplings of $\omega$ and $\rho$  mesons to describe baryon interacting in NSs.
We adopt the simplest NS model, assuming that the NS core consists of n, p, e and $\mu$ (npe$\mu$) matter.
Up to now, we do not known how the neutrino emissivity of the nucleonic direct URCA processes change if the tensor couplings of $\omega$ and $\rho$ mesons are included in NS matter.
The work mainly focus on the influences of the tensor couplings of $\omega$ and $\rho$ mesons on the maximum mass of NS, particle fraction and neutrino emissivity of the nucleonic direct URCA processes.

The effective Lagrangian density for the RHF approximation can be written as $\mathcal{L}= \mathcal{L}_0+ \mathcal{L}_I$, $\mathcal{L}_0$ is the Lagrangian density for the free nucleons, mesons and leptons,
\begin{equation}\label{1.1}
    \begin{split}
    \mathcal{L}_0=&\sum_N \overline{\Psi}_N[i \gamma_\mu \partial^\mu-m_{N}]\Psi_N+\frac{1}{2}(\partial_\mu \sigma \partial^\mu \sigma-m_\sigma^2 \sigma^2)-U(\sigma)\\
    &+\frac{1}{2}m_\omega^2 \omega_\mu \omega^\mu+\frac{1}{2}m_\rho^2 \bm{\rho}_\mu \bm{\rho}^\mu
    +\frac{1}{4}F_{\mu \nu} F^{\mu \nu}-\frac{1}{4}\bm{G}_{\mu \nu} \bm{G}^{\mu \nu}\\
    &+\sum_l \overline{\Psi}_l[i \gamma_\mu \partial^\mu-m_{l}]\Psi_l,\\
\end{split}
\end{equation}
where the potential function $U(\sigma)=\frac{1}{3}a\sigma^3+\frac{1}{4}b\sigma^4$, $F_{\mu\nu}=\partial_\mu \omega_\nu-\partial_\nu \omega_\mu$ and
$\bm{G}_{\mu \nu}=\partial_\mu \bm{\rho}_\nu-\partial_\nu\bm{\rho}_\mu$.
The interaction Lagrangian density $\mathcal{L}_I$ takes the form
\begin{equation}\label{1}
\begin{split}
    \mathcal{L}_I=&\sum_N \overline{\Psi}_N[g_{\sigma N}\sigma-g_{\omega N}\gamma_\mu \omega^\mu + \frac{f_{\omega N}}{2 M}\sigma_{\mu\nu} \partial^ \nu \omega^\mu\\
    &-g_{\rho N} \gamma_\mu {\bm{\rho}}^\mu \cdot \bm{I}_N+ \frac{f_{\rho N}}{2 M}\sigma_{\mu\nu} \partial^ \nu {\bm{\rho}}^\mu \cdot \bm{I}_N \\
    &-\frac{f_{\pi N}}{m_{\pi}} \gamma_5 \gamma_\mu \partial^ \mu \bm{\pi} \cdot \bm{I}_N ]\Psi_N,\\
\end{split}
\end{equation}
where the common scale mass M is the free nucleon mass, $\bm{I}_{N}$ is the isospin matrix of nucleon N. The
$g_{\sigma N}$, $g_{\omega N}$, $g_{\rho N}$ are the coupling constants of baryons to $\sigma$, $\omega$, $\rho$ mesons, respectively.
While $f_{\omega N}$, $f_{\rho N}$ and $f_{\pi N}$ are the isoscalar-, isovector-tensor and pseudovector coupling constants, respectively.

In the RHF approximation, the nucleon self-energy $\Sigma_{N}$ produced by the meson exchanges,
it can be written as

\begin{equation}\label{2.1}
    {\Sigma_{N}} (k)={\Sigma_{N}^S} (k)-{\gamma}_0 {\Sigma_N^0} (k) + \bm{ \gamma}{\cdot} \hat{\bm{k}}{\Sigma_N^V} (k),
\end{equation}
where $ \hat{\bm{k}}$ is the unit vector along $\bm{k}$. Using self-energy form (\ref{2.1}), the Dirac equation of nuclear matter can be written as
\begin{equation}\label{2.2}
    ({\gamma}_0 E^*_N-{\bm{ \gamma}} {\cdot} \bm{k}^*_N-M^*_N)\Psi_N=0,
\end{equation}
where $E^*_N$, $M^*_N$ and $k^*_N$ are defined by
\begin{equation}\label{2.3}
  E^*_N (k)=E_N(k)-\Sigma^0_N (k),
\end{equation}

\begin{equation}\label{2.4}
M^*_N (k)=M_N+\Sigma^S_N (k),
\end{equation}

\begin{equation}\label{2.5}
  \bm{ k}^*_N (k)=\bm{ k}_N+\bm{ \hat{k}} \Sigma^V_N (k),
\end{equation}
$E^*_N$ satisfies the relativistic mass-energy relation ${E^*_N}^2={\bm{ {k}}^*_N}^2+{M^*_N}^2$. $ E_N(k) $ is the single particle energy of nucleon in NS matter.
The nucleon self-energies ${\Sigma}_N^S (k)$, ${\Sigma}_N^0 (k)$ and ${\Sigma}_N^V (k)$ are given by\cite{Miyatsu2011Effects}
\begin{equation}\label{2.6}
\begin{split}
    {\Sigma}_N^S (k)=&-g_{\sigma N}\sigma_0+\sum \frac{I_{N {N}^{'}}^i}{(4 \pi)^2}
     \int_0^{p_{N^{'}}} dq q [\frac{M^*_{{N}^{'}} (q)}{E^*_{{N}^{'}} (q)}B_i (k,q)\\
    &+\frac{q^*_{{N}^{'}} (q)}{2 E^*_{{N}^{'}} (q)}D_i (k,q)],\\
\end{split}
\end{equation}

\begin{equation}\label{2.7}
    {\Sigma}_N^0 (k)=g_{\omega N } \omega_0+g_{\rho N } I_{N3} \rho_0+\sum \frac{I_{N {N}^{'}}^i}{(4 \pi)^2} \int_0^{p_{N^{'}}} dq q A_i (k,q),
\end{equation}

\begin{equation}\label{2.8}
\begin{split}
    {\Sigma}_N^V (k)=& \sum \frac{I_{N {N}^{'}}^i}{(4 \pi)^2} \int_0^{p_{N^{'}}} dq q [\frac{q^*_{{N}^{'}} (q)}{ E^*_{{N}^{'}} (q)}C_i (k,q)\\
    &+\frac{M^*_{{N}^{'}} (q)}{2E^*_{{N}^{'}} (q)}D_i (k,q)],\\
\end{split}
\end{equation}
where $p_{N}$ is the Fermi momentum of nucleon, $I_{N3}$ is the isospin projection of nucleon.
In the isoscalar ($\sigma$, $\omega$) and isovector channels ($\rho$, $\pi$), the isospin factors $I_{N {N}^{'}}^i$ are $\delta_{N {N}^{'}}$ and $2-\delta_{N {N}^{'}}$, respectively\cite{Sun2008Neutron}.

In Eq.\,(\ref{2.6}) and (\ref{2.7}), $\sigma_0$, $\omega_0$ and $\rho_0$ are the mean-field values of meson fields.
The meson field equations have the following form:
\begin{equation}\label{2}
    m_\sigma^2 \sigma_0+a \sigma_0^2+b \sigma_0^3=\sum_B \frac{g_{\sigma N}}{\pi^2} \int_0^{p_N} dk k^2  \frac{M^\ast_B(k) }{E^ \ast _N(k)},
\end{equation}
\begin{equation}\label{3}
    m_\omega^2 \omega_0 =\sum_B g_{\omega N} \frac{p_N^3}{3\pi^2},
\end{equation}

\begin{equation}\label{4}
    m_\rho^2 \rho_0 =\sum_N g_{\rho N} I_{3N} \frac{p_N^3}{3\pi^2}.
\end{equation}
The nucleon density is expressed by
\begin{equation}\label{4.1}
    \rho_N=\frac{p_N^3}{3\pi^2}.
\end{equation}

The chemical potentials of nucleons and leptons satisfy the following relationship under $\beta$ equilibrium conditions,
\begin{equation}\label{8}
    \mu_p=\mu_n-\mu_e,\quad  \mu_\mu=\mu_e.
\end{equation}
At zero temperature the chemical potentials of nucleons and leptons are expressed by
\begin{equation}\label{8.1}
    \mu_N=E^*_N (p_N)+{\Sigma}_N^0  (p_N),
\end{equation}

\begin{equation}\label{9.1}
    \mu_l=\sqrt{p_l^2+{m_l^2}}.
\end{equation}
The nucleons and leptons should meet the charge neutrality and baryon number conservation conditions, which are given by
\begin{equation}\label{7}
    \sum_N q_N \rho_N-\rho_e-\rho_\mu=0,
\end{equation}
\begin{equation}\label{7.1}
    \sum_N \rho_N=\rho,
\end{equation}
where $\rho$ is the total nucleonic density.

The total energy density of NS matter is written by\cite{Miyatsu2011Effects}
\begin{equation}\label{7.2}
\begin{split}
    \varepsilon=& \sum_N \frac{1}{\pi^2}\int_0^{p_{N}} dk k^2 [T_N (k)+\frac{1}{2}V_N (k) ] -\frac{\sigma_0^3}{2}(\frac{a}{3}+\frac{b}{2}\sigma_0) \\
    &+\sum_l  \frac{1}{\pi^2} \int_0^{p_{l}} dk k^2  \sqrt{k^2+m_l^2}, \\
    \end{split}
\end{equation}
where $T_N (k)$ and $V_N (k)$ are defined as
\begin{equation}\label{7.3}
     T_N (k)= \frac{ M_N M_N^*(k)+k k_N^*(k)}{E_N^*(k)},
\end{equation}

\begin{equation}\label{7.4}
    V_N (k)=\frac{ M_N^* \Sigma^s_N (k)+  k_N^* \Sigma^v_N (k)}{E_N^*(k)}+\Sigma^0_N (k).
\end{equation}
The pressure of NS matter can be obtained by

\begin{equation}\label{7.5}
    P=\rho^2\frac{\partial}{\partial\rho}(\frac{\varepsilon}{\rho}).
\end{equation}

The simplest possible neutrino emission processes are composed of two successive
reactions, beta decay and capture:
\begin{equation}\label{9}
    N_1 \rightarrow {N}_2+l+{\overline{\nu}}_l, N_2+l \rightarrow {N}_1+{\nu}_l,
\end{equation}
where $N_1$ and $N_2$ are nucleons,
and $l$ is a lepton. The relativistic neutrino emissivity of the direct URCA processes among nucleons and electrons can be written by
given by the Fermi golden rule, expressed as follows:
\begin{equation}\label{10}
\begin{split}
    Q_R=& \frac{457\pi}{10080}G_F^2 C^2 T^6 \Theta (p_e+p_{N_2}-p_{N_1}) \\
    & \times \{ f_1 g_1[(\varepsilon_{N_1}+\varepsilon_{N_2})p_l^2-(\varepsilon_{N_1}-\varepsilon_{N_2})(p^2_{N_1}-p^2_{N_2})]\\
    &+2g_1^2 \mu_e m^\ast_{N_1} m^\ast_{N_2}+(f^2_1+g^2_1)[\mu_e(2\varepsilon_{N_1}\varepsilon_{N_2}-m^\ast_{N_1} \ast\\ &m^\ast_{N_2})
    +\varepsilon_{N_1}P_l^2-\frac{1}{2}(p^2_{N_1}-p^2_{N_2}+p^2_e)(\varepsilon_{N_1}+\varepsilon_{N_2})],\\
\end{split}
\end{equation}
where $G_F=1.436 \times 10^{-49}$ erg $cm^3$ is the weak-coupling constant, $f_1$, $g_1$ and $C$ are the vector, axial-vector constants,
and Cabibbo factor, respectively\cite{Prakash1992Rapid}. $\mu_e$ is the chemical potential of a electron.
$\varepsilon_{N_1}$ and $\varepsilon_{N_2}$ are the nucleonic kinetic energy.
$\Theta(x)=1$ if $x\geq0$ and zero otherwise.

Solving the above equations consistently, we can acquire a series of physical quantities, like the Fermi momentum, effective mass, particle fraction and chemical potential of nucleons and electrons at a given total nucleonic density $\rho$.
They are taken as inputs and by which is used to solve Eq.\,(25). The relativistic neutrino emissivity for the nucleonic direct URCA processes can be obtained.

In the paper, we mainly consider the influence of the tensor couplings of $\omega$ and $\rho$ mesons on the nucleonic direct URCA processes in npe$\mu$ matter. According to whether including the tensor couplings of $\omega$ and $\rho$ mesons or not, the following discussion analyses the four cases in detail. A1 is not inclusive for the tensor couplings of $\omega$ and $\rho$ mesons in NS matter. Instead, A2 includes the tensor couplings of $\omega$ and $\rho$ mesons in NS matter.
B1 also does not consider the tensor couplings of $\omega$ and $\rho$ mesons in NS matter. B2 is the opposite of B1, which only considers the tensor coupling of $\rho$ meson in NS matter. The bulk properties of NSs are obtained for the four cases with the parameter sets in Table 1.

\noindent{\footnotesize Table 1.Parameter sets\cite{Katayama2012EQUATION,H1997Neutron}.

\vskip 2mm \tabcolsep 6pt

\centerline{\footnotesize
\begin{tabular}{ccccc}\hline
Parameter & $g_{\sigma N} $  & $g_{\omega N}$ & $f_{\omega N}$& $g_{\rho N}$ \\
\hline
A1(no tensor)  &  7.66061  &  7.0096  & 0& 2.33653  \\
A2(with tensor)  &   6.89209  &   7.2908  & -0.8749 &  2.43027 \\
B1(no tensor) & 9.28353 & 8.37378 & 0& 2.10082 \\
B2(with tensor) & 9.24264 & 8.25548 & 0& 2.19809 \\
\hline   & $f_{\rho N}$ &$f_{\pi N}$ & $a$&$b$\\
\hline
A1(no tensor)   & 0& 1.00265 &10.7 &170 \\
A2(with tensor)  &  8.99199& 1.00265 & 10.5 &235 \\
B1(no tensor)   & 0& 1.00265 &12.69240 & -15.98724 \\
B2(with tensor)   & 8.13293 & 1.00265 & 11.14070 & -19.60256 \\
\hline
\end{tabular}}}

\vskip 2mm

\noindent{\footnotesize Table 2. Maximum masses $M_{max}$, and
corresponding radii R and center densities $\rho_c$ for the four cases..

\vskip 2mm \tabcolsep 6pt

\centerline{\footnotesize
\begin{tabular}{cccc}\hline
 Parameter & $M_{max}(M_\odot) $  &R(km) & $\rho_c$(fm$^{-3}$) \\
\hline
A1 & 1.99   & 11.06   & 1.07  \\
A2 &  2.00   & 10.66   & 1.10  \\
B1 & 2.31 & 10.98 & 0.98   \\
B2 & 2.33 & 10.91 & 0.98    \\
\hline
\end{tabular}}}

\vskip 0.5\baselineskip

The mass-radius relations of NSs are obtained by substituting EOS into the Tolman-Oppenheimer-Volkoff (TOV) equations\cite{Tolman1939Static,Oppenheimer1939On}.
They are shown in Fig.\,1 for the four cases.
The maximum masses $M_{max}$, the corresponding radii R and center densities $\rho_c$ for the four cases are listed in Table 2.
As seen in Fig.\,1 and Table 2, the inclusion of the tensor couplings of $\omega$ and $\rho$ mesons or the tensor coupling of $\rho$ meson increases the maximum mass
of NSs.
The maximum mass of NSs for the four cases are agreement with the observed values of PSR J 1614-2230.
Fig.\,2 shows the particle fraction $Y_i$ as a function of the total nucleonic density $\rho$ for the four cases.
When the tensor couplings of $\omega$ and $\rho$ mesons are included(or the tensor coupling of $\rho$ meson), namely the case A2(or case B2), the neutron fraction $Y_n$ decreases and proton fraction increases.
The increment of the proton fraction would result in an increment of the lepton fraction due to the charge neutrality and $\beta$ equilibrium conditions.
Moreover, the inclusion of the tensor coupling of $\rho$ meson(case B2) makes the conclusion from above more obviously than the tensor couplings of $\omega$ and $\rho$ mesons(case A2).
According to Eq.\,(14), these changes of the individual Fermi momenta of neutrons and protons can be obtained for the four cases.
Fig.\,3 represents the effective masses of nucleons as a function of the total nucleonic density $\rho$ for the four cases.
The effective masses of nucleons for the case A2 and B2 are lower than the corresponding values for the case A1 and B1 firstly and then are higher than the corresponding values for the case A1 and B1, respectively.
These changes must change the neutrino emissivity of the nucleonic direct URCA processes.

\begin{figure}
\includegraphics[width=8.5cm]{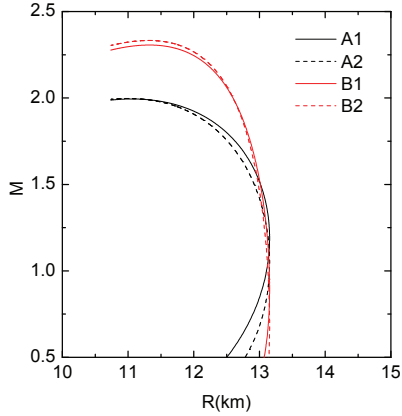}
\caption{\label{fig1}The mass-radius relations of NS for the four cases.}
\end{figure}

\begin{figure}
\includegraphics[width=8.5cm]{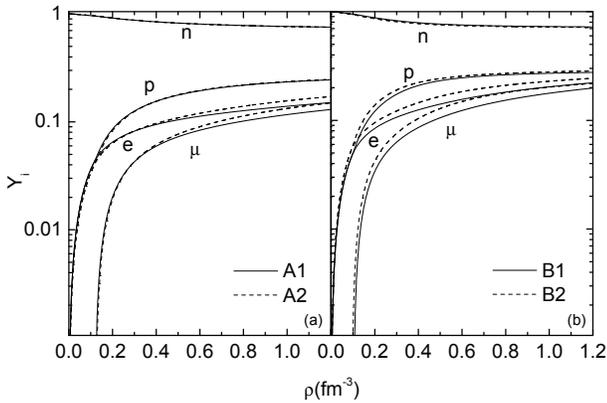}
\caption{\label{fig2}The particle fractions of baryons and
leptons as a function of the total nucleonic density $\rho$ for the four cases in npe$\mu$ matter.}
\end{figure}

\begin{figure}
\includegraphics[width=8.5cm]{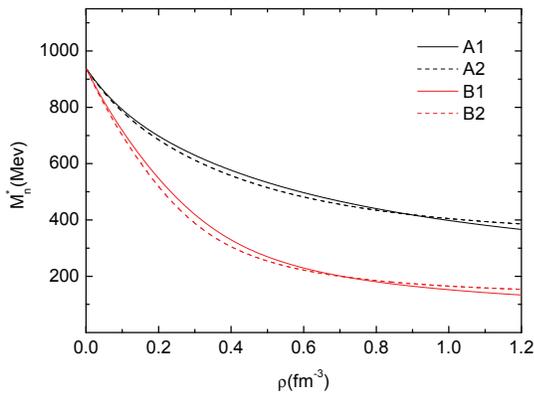}
\caption{\label{fig3}The nucleonic effective masses as a function
of the total nucleonic density $\rho$ for the four cases in npe$\mu$ matter.}
\end{figure}

The relativistic neutrino emissivity of the direct URCA processes among nucleons and electrons as a function of the total nucleonic density $\rho$ for the four cases are plotted in Fig.\,4.
The nucleonic direct URCA processes occur when the triangle condition $p_{p}+p_{e}>p_{n}$ in Eq.\,(25) is satisfied.
We can see that the nucleonic direct URCA processes for the cases A1 and A2 happen at the same time.
While the threshold dendity of the nucleonic direct URCA processes for the case B2 is lower than the corresponding values for the  case B1.
In addition, the relativistic neutrino emissivity $Q_R$ for the case A2 is obviously larger than the corresponding values for the case A1.
However, the relativistic neutrino emissivity $Q_R$ in the case B2 is less than the corresponding values in the case B1 firstly and then increases, equals or exceeds the values in case B1.
It may be concluded that the inclusion of the tensor couplings of $\omega$ and $\rho$ mesons would accelerate the nonsuperfluid
NS cooling in the mass range of happening the nucleonic direct URCA processes.
While the inclusion of the tensor coupling of $\rho$ meson, we could conclude that the cooling rate becomes slow for a low-mass NS and rapid for a massive NS.

\begin{figure}
\includegraphics[width=8.5cm]{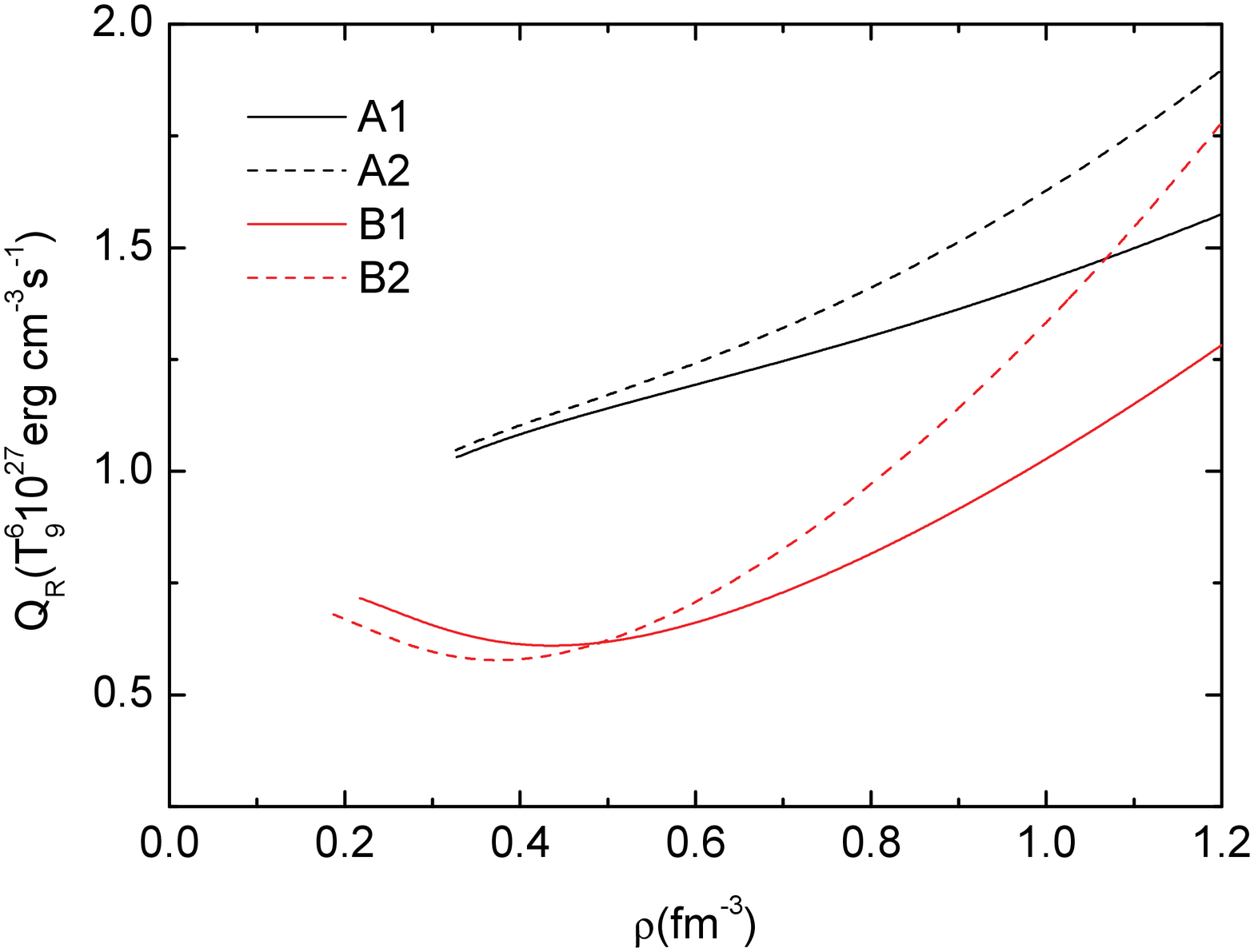}
\caption{\label{fig4}The relativistic neutrino emissivity $Q_{R}$
as a function of the total nucleonic density $\rho$ for the four cases in npe$\mu$ matter.}
\end{figure}

In summary, we have been investigated the influences of the tensor couplings of $\omega$ and $\rho$ mesons on the nucleonic direct URCA processes by adopting the RHF theory in npe$\mu$ matter.
We find that the inclusion of the tensor couplings of $\omega$ and $\rho$ mesons(or the tensor coupling of $\rho$ meson) leads to a small increase of the NS maximum mass.
In addition, it makes an obvious enhancement of the relativistic neutrino emissivity $Q_{R}$ for the nucleonic direct URCA processes.
Which means that the inclusion of the tensor couplings of $\omega$ and $\rho$ mesons would quicken up the cooling rate of the non-superfluid NS.
However, when we only consider the tensor coupling of $\rho$ meson in NS matter, the relativistic neutrino emissivity $Q_{R}$ for the nucleonic direct URCA processes decreases at low densities and increases at high densities.
In other word, the tensor coupling of $\rho$ meson causes the NS cool to slow for a low-mass NS and speed up for a massive NS.

\bibliography{sample1}

\end{document}